\documentclass[10pt]{iopart}
\pdfoutput=1
\usepackage{fullpage}
\usepackage{iopams,graphicx,color}
\usepackage{setstack}
\usepackage[colorlinks=true]{hyperref} 
\hypersetup{urlcolor=blue,linkcolor=blue,citecolor=blue,colorlinks=true} 
\usepackage[dvipsnames]{xcolor}

\usepackage{color}
\usepackage{pgf,tikz}
\usepackage{verbatim}
\usepackage{marginnote}
\usepackage{enumitem}

\usepackage{braket}

\begin{document}
 \title{Phase transitions in distributed control systems with multiplicative noise}

\author{Nicolas Allegra$^{1,2}$,
Bassam Bamieh$^1$, Partha Mitra$^3$ and Cl\'ement Sire$^{4}$}

\address{
$^1$ Department of Mechanical Engineering, University of California, Santa Barbara, USA  \\
$^3$ Cold Spring Harbor Laboratory, Cold Spring Harbor, NY, USA\\
$^4$ Laboratoire de Physique Th\'eorique, Universit\'e de Toulouse, UPS,
CNRS, F-31062 Toulouse, France\\}
 
\begin{abstract}
\textcolor{black}{
Contemporary technological challenges often involve many degrees of freedom in a distributed or networked setting. Three aspects are notable: the variables are usually associated with the nodes of a graph with limited communication resources, hindering centralized control; the communication is subjected to noise; and the number of variables can be very large. These three aspects make tools and techniques from statistical physics particularly suitable for the performance analysis of such networked systems in the limit of many variables (analogous to the thermodynamic limit in statistical physics). Perhaps not surprisingly, phase-transition like phenomena appear in these systems, where a sharp change in performance can be observed with a smooth parameter variation, with the change becoming discontinuous or singular in the limit of infinite system size. \\
In this paper we analyze the so called network consensus problem, prototypical of the above considerations, that has been previously analyzed mostly in the context of additive noise. We show that qualitatively new phase-transition like phenomena appear for this problem in the presence of multiplicative noise. Depending on dimensions and on the presence or absence of a conservation law, the system performance shows a discontinuous change at a threshold value of the multiplicative noise strength. In the absence of the conservation law, and for graph spectral dimension less than two, the multiplicative noise threshold (the stability margin of the control problem) is zero. This is reminiscent of the absence of robust controllers for certain classes of centralized control problems. Although our study involves a "toy" model we believe that the qualitative features are generic, with implication for the robust stability of distributed control systems, as well as the effect of roundoff errors and communication noise on distributed algorithms. } 

\end{abstract}
 \date{\today}
\pacs{05.10.-a, 02.30.Yy
, 89.75.Fb}
\submitto{JSM}
\maketitle

\tableofcontents

\newpage

\section{Introduction}

Phase transition phenomena have played a central role in twentieth
century theoretical physics, ranging from condensed matter physics to
particle physics and cosmology. In recent decades, phase transition
phenomena, corresponding to non-analytic behavior arising from large
system-size limit in systems of many interacting variables, have
increasingly appeared in the engineering disciplines, in
communications and computation \cite{mezard2009information},
robotics \cite{olfati2006flocking,chowdhury2000statistical}, control
theory \cite{olfati2007consensus,liu2011controllability} and machine
learning. Related phenomena have also been noted in behavioral
biology \cite{vicsek2012collective}, social sciences
\cite{castellano2009statistical} and economics
\cite{bouchaud2003theory,mantegna1999introduction}.  It is not as
widely appreciated that phase transition-like behavior may also be
observed in distributed control systems (or distributed algorithms)
with many variables, for similar mathematical reasons that they
appear in statistical physics, namely the presence of many
interacting degrees of freedom.

Problems in a variety of engineering areas that involve
interconnections of dynamic systems are closely related to consensus
problems for multi-agent systems \cite{lns-v.85}. The problem of
synchronization of coupled oscillators \cite{kuramoto2012chemical}
has attracted numerous scientists from diverse fields
(\cite{acebron2005kuramoto} for a review). Flocks of mobile agents
equipped with sensing and communication devices can serve as mobile
sensor networks for massive distributed sensing
\cite{cortes2002coverage,olfati2006flocking,olfati2007consensus}. In
recent years, network design problems for achieving faster consensus
algorithms has attracted considerable attention from a number of
researchers \cite{xiao2004fast}. Another common form of consensus
problems is rendezvous in space \cite{lin2004multi,lin2007multi}.
This is equivalent to reaching a consensus in position by a number
of agents with an interaction topology that is position induced.
Multi-vehicle systems are also an important category of networked
systems due to their technological applications
\cite{ren2008distributed}. Recently, consensus algorithms had been generalized to quantum systems \cite{mazzarella2015consensus} and opening new research directions towards distributed quantum information applications \cite{nielsen2010quantum}.

In contrast to the above mentioned work, a relatively less studied problem is that of robustness or resilience of the algorithms 
to uncertainties in models, environments or interaction signals. This is the central question in the area of Robust Control, but it 
has been relatively less studied in the context {\em distributed} control systems. This issue is quite significant since algorithms that 
work well for a small number of interacting subsystems may become arbitrarily fragile in the large-scale limit. Thus, of
 particular interest
to us in this paper is how large-scale distributed control
algorithms behave in various uncertainty scenarios. Additive noise
models have been studied in the context of networked consensus
algorithms~\cite{xiao2007distributed,huang2009coordination},
including scaling limits for large
networks~\cite{patterson2014consensus,bamieh2012coherence}. More
relevant to the present work are uncertainty models where link or
node failures,  message distortions, or randomized
algorithms~\cite{fagnani2008randomized,carli2010gossip,patterson2010convergence,wang2012distributed,wang2010consensus}
are modeled by multiplicative noise. In this latter case,  the
phenomenology is much richer than the case of only additive noise. The basic building block of multi-agent and distributed control systems is 
the so-called consensus algorithm. 
In this paper we develop a rather general model of consensus
algorithms in random environments that covers the above mentioned
uncertainty scenarios, and we study its large-size scaling limits
for $d$-dimensional lattices.
We now motivate the problem formulation of this paper using a simple version of the consensus algorithm. 

\subsection{Example of a simple Laplacian consensus algorithm}

We consider a local, linear first-order consensus algorithms over an undirected, connected network modeled by an undirected graph $G$ with $N$ nodes and $M$ edges. We denote the adjacency matrix of $G$ by $A$, and $D$ is the degree matrix. The Laplacian matrix of the graph $G$ is denoted by $L$ and is defined as $L=D-A$. In the first-order consensus problem, each node $i$ has a single state $u_i(t)$. The state of the entire system at time $t$ is given by the vector ${\bf u}(t)\in\mathbb{R}^N$ . Each node state is subject to stochastic disturbances, and the objective is for the nodes to maintain consensus at the average of their current states. The dynamics of this system in continuous time is given by 
\begin{equation}\label{consensus}
\dot{{\bf u}}(t)=-\alpha L {\bf u}(t)+{\bf n}(t),
\end{equation}
where $\alpha$ is the gain on the communication links, and ${\bf n}(t)\in\mathbb{R}^N$ is a white noise. 
This model covers a large number of systems. A few examples are 
\begin{itemize} 
	\item In flocking problems $u_i(t)$ is the current heading angle of a vehicle or agent \cite{vicsek2012collective}.
			$L$ is the Laplacian of the agents'
			connectivity network determined by e.g. a distance criterion. The disturbance $n_i(t)$ models random forcing on 
			the $i$'th agent. 
	\item In load balancing algorithms for a distributed computation network $u_i(t)$ is the current load on a computing node. 
			$L$ is the Laplacian of the connectivity graph \cite{cybenko1989dynamic,boillat1990load} . 
			The disturbance $n_i(t)$ models arriving (when positive) or completed (when negative) jobs. 
	\item In distributed sensing networks \cite{cortes2002coverage,olfati2006flocking,olfati2007consensus}, the $i$'th sensor makes a local measurement $u_i(0)$ of a global quantity. The aim is for the 
			sensors to communicate and agree on the value of the sensed quantity without having a central authority, thus they 
			communicate locally based on a connectivity graph with $L$ as its Laplacian. The dynamics~(\ref{consensus}) represents 
			averaging of each sensor with their neighbors, while the disturbances $n_i(t)$ can represent the effective noise in 
			communicating with neighbors. The aim is for all sensors to reach the same estimate (which would be the initial network 
			mean) asymptotically, i.e. for each $i$, $\lim_{t\rightarrow\infty} u_i(t) ~=~ \frac{1}{N} \sum_i u_i(0)$. 
\end{itemize}

We should note that in much of the literature, the algorithm Eq.~\eref{consensus} has been studied for finite systems and in the absence
of disturbances ${\bf n}(t)$. In the large-scale limit however, there are significant differences in phenomenology
 between the disturbance-free versus  the uncertain scenarios. 
In the absence of disturbances and for a connected graph, the system Eq.~\eref{consensus} converges asymptotically to the average of the initial states.
With the additive noise term, the nodes do not converge to consensus, but instead, node values fluctuate around the average of the current node states. Let us denote the spatial average across the network ("network mean") of the sites by $m(t)$,   
 
 \begin{equation}\label{meanM}
m(t)=\frac{1}{N} \sum_{k=1}^{N}u_k(t).
\end{equation}

In the absence of noise, $m(t)=m(0)$ for a finite graph. Although $m(t)$ shows a diffusive behavior for finite sized networks in the presence of additive noise, the diffusion coefficient is proportional to $1/N$ so that in the infinite lattice size limit one still obtains $m(t)=m(0)$. This can be seen by spatially averaging the consensus equation. The noise-free dynamics preserves the spatial average (equivalently, the Graph Laplacian is chosen to be left stochastic), so in the absence of noise $ \dot{m}(t) = 0$. In the presence of the additive noise, 

\begin{equation}\label{consensus}
\dot{m_N}(t)=\frac{1}{N} \sum_i n_i(t),
\end{equation}
 
Assuming the noise is uncorrelated between sites, this means that $var(m_N(t)) \propto \frac{t}{N}$, so that $\lim_{N\rightarrow \infty} var(m_N(t)) = 0$. Thus, in the infinite lattice size limit, $m(t)=m(0)$. Note that this continues to be the case for the multiplicative noise model discussed in the paper when there is a conservation law that preserves the network mean in the absence of the additive noise component. The spatial variance across the lattice sites has been called the "network coherence" in previous work \cite{bamieh2012coherence}. 

\begin{equation}\label{coherence}
C^{\infty}_N:=\lim_{t\rightarrow\infty} \frac{1}{N}\sum_{i=1}^{N} \mathrm{var}\Big(u_i(t)- m(t)  \Big),
\end{equation}

Note that in the infinite size limit for the consensus problem with additive noise $m(t)=m(0)$ so the spatial variance ("network coherence") also characterizes the variance of the individual site variables from the desired mean $m(0)$ at the initial time. If this variance is finite in the limit of large N, then under the consensus dynamics the individual lattice sites settle down to stationary fluctuations around the desired initial spatial mean, and this quantity can be recovered from a single site at long times, by taking a time average. 

It has been shown that $C^{\infty}_N$ is completely determined by the spectrum of the matrix $L$ (see \cite{lns-v.85} for a review). Let the eigenvalues of $L$ denoted by $0=\lambda_1<...<\lambda_N$. The network coherence is then equal to 
\begin{equation}\label{EV}
C^{\infty}_N:=\frac{1}{2\alpha N}\sum_{i=2}^{N}  \frac{1}{\lambda_i}.
\end{equation}
A classical result \cite{bamieh2012coherence} shows that, if one considers the network to be $\mathbb{Z}^d$, then the large-scale ($N\rightarrow\infty$) properties of the coherence show the following behavior
\begin{eqnarray}\label{scalingC}
C^{\infty}_N \sim \left\{
    \begin{array}{ll}
        N & \mbox{in} \ d=1 \\
        \log{N} & \mbox{in} \ d=2 \\
        1 & \mbox{in} \ d>2.
    \end{array}
\right.
\end{eqnarray}
Compared to a perfect Laplacian algorithm (without additive noise), the addition of disturbances makes the system unable to reach the global average in low dimensions. In higher dimensions, the network coherence becomes finite and the algorithm performs a statistical consensus at large time. 
This result shows that there are fundamental limitations to the efficiency of the algorithm for a large-scale system in presence of additive noise in low dimensions. Let us mention that this algorithm can be extended to a second-order dynamic where each nodes has two state variables $u(t)$ and $v(t)$. These system dynamics arise in the problem of autonomous vehicle formation control (the state of a node is given by the position and velocity of the vehicle). The vehicles attempt to maintain a specified formation traveling at a fixed velocity while subject to stochastic external perturbations and a similar coherence definition can be introduced, and the scaling of the coherence with $N$ has been computed \cite{bamieh2012coherence} and similar dimensional limitations has been unraveled.
\\
\\

\subsection*{Robustness and Multiplicative Noise} 

In engineering systems, $N$ may be large but perhaps not as large as might be considered in condensed matter problems. 
In the distributed control setting, one can give a robustness interpretation 
of asymptotic limits like Eq.~\eref{scalingC}. For example, in the $d=1$ case (which is relevant to the so-called automated vehicular platoons
problem~\cite{bamieh2012coherence}) one can say that as the formation size increases, it becomes susceptible to arbitrarily small forcing
disturbances. A more proper robustness interpretation however requires considering scenarios with {\em multiplicative} noise as a model 
for uncertain system dynamics, and not just uncertain forcing or measurement noise (which is typically modeled with {\em additive} noise). 
Consider the following more general version of Eq.~\eref{consensus}
\begin{equation}\label{consensusTV}
\dot{{\bf u}}(t)=-\alpha L(t) ~ {\bf u}(t)+{\bf n}(t),
\end{equation}
where the Laplacian $L(t)$ is now a time-varying matrix with some ``structured'' randomness. This is now a multiplicative noise model 
since the randomness multiplies the state ${\bf u}(t)$. 
This is used to model effects of 
random environments such as  
networked algorithms in which links or nodes may fail randomly, randomly varying system parameters,  round-off error in floating-point arithmetic, or more generally any 
random errors that are proportional to the current state~\cite{fagnani2008randomized,carli2010gossip,patterson2010convergence,wang2012distributed,wang2010consensus}. If arbitrarily small probabilities of error in $L(t)$ can produce large effects
in the dynamics of Eq.~\eref{consensusTV} then the algorithms are fragile and lack robustness. 


The main question in this paper is to study a fairly general model of networked algorithms in random environments like Eq.~\eref{consensusTV}. In particular, we study scaling limits in $d$-dimensional lattices, and
 characterize how the system  responds to multiplicative noise and whether (or not) the algorithm can remain stable and perform its task once some arbitrary small multiplicative noise in incorporated. It is highly unlikely that taking account of multiplicative noise (with variance $\sigma^2$) is going to improve the scaling in low dimensions, but the challenge is to characterize the robustness of the algorithm in high dimensions. This measure of robustness can be quantified by the amount of noise variance $\sigma^2$ that the system can sustain before hitting a threshold $\sigma^2_c$ above which the algorithm becomes unstable and the coherence grows unboundedly. The range of noise 
\begin{equation}\label{stabilitymargin}
0\leqslant\sigma^2<\sigma^2_c, 
\end{equation}
in which the coherence is bounded is often called margin of stability in the control literature \cite{doyle2013feedback}. 
The main goal of this article is to quantify this margin, and to explicitly give its behavior in a wide range of models.

\subsection{Plan of the article}

The plan of the article is the following, after the introduction of
the general system and its main properties, we shall discuss the several
conservation laws that one might consider and the effects on the
correlations in the system. Furthermore, different cases of the form
of the random environment, which is modeled by multiplicative random variables
will be analyzed, leading to various types of correlations between
the different degrees of freedom of our model. The main part of the
article will be the study of the time-behavior of the network coherence Eq.~\eref{timecoherence} {\it via} the 2-point correlation
function. We shall see that, depending on the conservation law that
we are imposing in our system, different behaviors of the coherence is found. The stability margin Eq.~\eref{stabilitymargin}, introduced in the previous section, shall be fully characterized and its explicit value will be given for each considered cases.
The complete phase diagram of the network-coherence will be explored in any dimension, with an emphasis on the presence of a noise- threshold between various large-scale behaviors. The phase diagrams shall be explained in a more general point of view by analyzing the various universality classes of the system and making contact with the well-known phenomenology of disordered systems. Although our results are derived for first-order consensus algorithms Eq.~\eref{consensus}, we expect qualitatively similar behaviors to occur in higher-order consensus algorithms such as those used in vehicular formation control \cite{bamieh2012coherence}.

\section{Consensus algorithms in a random environment}

\subsection{Definition of the system and conservation laws}

In this section, one introduces a class of algorithms generalizing the noisy Laplacian algorithm Eq.~\eref{consensus}. The generalized algorithm is a discrete-time stochastic evolution for the quantity $u_i(t)\in{\mathbb R}$ with both additive and multiplicative noises. Let us define the system with a general coupling kernel
${\mathrm K}$ on a square-lattice of size $L$ and dimension $d$ 
\begin{equation}\label{eq1}
u_i(t+1)=\sum_{j}\xi_{ij}(t){\mathrm K}({\bf r}_i,{\bf r}_j)u_j(t)+V_i(t)u_i(t)+n_i(t),
\end{equation}
where $\xi_{ij}(t)$, $V_i(t)$ are random variables with mean
$\langle\xi\rangle$ and $\langle V\rangle$, and ${\mathcal V_i}$
being the neighborhood of the site $i$. Here we choose the
convention that $\xi_{ii}(t)=0$ such that the off-diagonal randomness
is only in $V_i$. The link $\xi_{ij}(t)$ and onsite $V_{i}(t)$
variables are Gaussian variables with correlations that will be detailed in the next sections.
The additive noise is centered and uncorrelated in space and time with variance $\sigma_n^2$
\begin{eqnarray}
&&\langle n_i(t)\rangle=0,\\
&&\langle n_i(t)n_j(t')\rangle=\sigma_n^2 \delta_{ij}\delta_{tt'},
\end{eqnarray}
where $\delta_{ab}$ is the Kronecker symbol. The system is translationally invariant so that ${\mathrm K}({\bf
r}_i,{\bf r}_j)={\mathrm K}({\bf r}_i-{\bf r}_j)$. In the following,
we shall use the convenient notation ${\mathrm K}_{i-j}:={\mathrm
K}({\bf r}_i-{\bf r}_j)$. 
We assume also that the distributions do not vary with time. 
As a consequence, all the moments of the different variables do not depend of the space-time
coordinates ( {\it e.g.} $\langle \xi_{ij}(t)^p \rangle
=\langle\xi^p\rangle$  $\forall i,j$). Let us define the variances $\sigma_{\xi}^2=\langle \xi^2 \rangle-\langle \xi\rangle^2
$ and $\sigma_{V}^2=\langle V^2 \rangle-\langle
V\rangle^2$.
In the spirit of Eq.~\eref{consensus}, the system can be written in a matrix form
\begin{equation}\label{matrixform}
{\bf u}(t+1)= {\mathrm M}(t) {\bf u}(t)+{\bf n}(t),
\end{equation}
 where the matrix ${\mathrm M}(t)$ is a time-dependent random matrix with matrix elements ${\mathrm M}_{ij}(t)={\mathrm
 K}_{i-j}\xi_{ij}(t)+\delta_{ij}V_j(t)$ and ${\bf n}(t)$ is the additive noise vector.  This algorithm is a generalization of Eq.~\eref{consensus} in which the Laplacian matrix is replaced by a time-dependent random matrix.  
 The problem is then related to the asymptotic properties of product of random matrices  (see
 \cite{crisanti2012products} or \cite{touri2012product,muller2002asymptotic} for example in the
 context of consensus and wireless communications).  The exponential
 growth rate of the matrix powers ${\mathrm M}^t$ as
 $t\rightarrow\infty$ is controlled by the eigenvalue of ${\mathrm
 M}$ with the largest absolute value. While the stationary distribution was guaranteed in the
additive model Eq.~\eref{consensus}, the multiplicative model Eq.~(\ref{eq1}) may or may
not have a stationary solution depending on the parameters and the
dimension $d$.  The mean value can be written
\begin{equation}\label{meanvalue}
\langle u_i(t+1)\rangle=\big(\|{\mathrm K}\|_{1}\langle
\xi\rangle+\langle V\rangle\big) \langle u_i(t)\rangle,
\end{equation}
where $\|{\mathrm K}\|_{1}=\sum_i {\mathrm K}_i$.  Then obviously
$\langle u_i(t)\rangle=0$ when $\langle V\rangle=-\|{\mathrm
K}\|_{1}\langle \xi\rangle$ and $\langle u(t+1)\rangle=\langle
u_i(t)\rangle$ when $\langle V\rangle=1-\|{\mathrm K}\|_{1}\langle
\xi\rangle$. In general, the sequence converges if $|(\|{\mathrm
K}\|_{1}\langle \xi\rangle+\langle V\rangle | \leqslant 1$ and
diverges otherwise for a positive initial condition $\langle
u(0)\rangle>0 $. 
One way of studying the margin will be to compute the time-behavior of the coherence
\begin{eqnarray}\label{timecoherence}
C_L(t)=\frac{1}{L^d}\sum_{r=1}^{L^d}\langle (u_r(t)-m(t))(u_0(t)-m(t)))\rangle=\frac{1}{L^d}\sum_{r=1}^{L^d}{\mathrm G}_{r}(t)-\langle m(t)\rangle^2,
\end{eqnarray}
where ${\mathrm G}_{r}(t)$ is the local 2-point correlation function $\langle u_r(t)u_0(t)\rangle$ that captures correlations between an arbitrary node $0$ and a node $r$ and $m(t)=L^{-d}\sum_{i}u_i(t)$ being the space average of the value $u_i(t)$. The term $\langle m(t)\rangle^2$ is stationary if  $|(\|{\mathrm K}\|_{1}\langle \xi\rangle+\langle V\rangle | \leqslant 1$. Therefore the time-evolution of $C_L(t)$ is simply governed by the time-evolution of ${\mathrm G}_{r}(t)$. This function will be the main quantity that one will be interested in to characterize the scaling behavior of the performance of the algorithm. 
In the following, we shall impose the time conservation of the average $m(t)$ either exactly or in average at the thermodynamic limit $L\rightarrow\infty$.

\subsection{Correlations induced by exact conservation}

In this section, we summarize all the particular cases of
Eq.~\eref{matrixform}, while preserving conservation of $m(t)$
exactly ({\it i.e} $\sum_{i}{\mathrm M}_{ij}=1$) at the thermodynamic limit. The randomness is
fully defined by specifying the form of the correlations $\langle{\mathrm
M}_{ij}(t){\mathrm M}_{pq}(t)\rangle$.
The exact conservation is defined by $m(t+1)=m(t)$ . As a consequence the matrix ${\mathrm
 M}$ satisfies column-stochasticity $\sum_{i}{\mathrm
 M}_{ij}=1$. Hence, it is straitforward to show that we need to impose
\begin{equation}\label{xiV}
V_j(t)=1-\sum_{i}\xi_{ij}(t){\mathrm K}_{i-j},
\end{equation}
 for that conservation law to be satisfied Let us notice that the summation is over the first
index, while it is on the second in the evolution equation Eq.~\eref{eq1}.  In this case, the multiplicative quantities are linearly dependent and one can write everything in terms of the variance of one or the other.
 Additionally one can assume many different form for the link variables $\xi_{ij}(t)$. The simplest
case is to assume that $\xi_{ij}(t)$ and $\xi_{ji}(t)$ are
independent, this assumption does not add any new correlation
between the onsite variables $V_i(t)$. The opposite scenario is to consider
$\xi_{ij}(t)=\xi_{ji}(t)$ which enforces correlations between
onsite variables as we shall see in the following. We will also
consider both cases where the link variables $\xi_{ij}(t)$, depend
only on one index, {\it i.e.} $\xi_{ij}(t)=J_j(t)$ and $\xi_{ij}(t)=g_i(t)$ which
we call respectively isotropic and gain noise cases.  For each considered possibilities, we shall explicitly
compute the relevant correlations that we will need later on for the calculation of the coherence and in particular for the calculation of ${\mathrm G}_r(t)$.

\subsubsection{Asymmetric links}

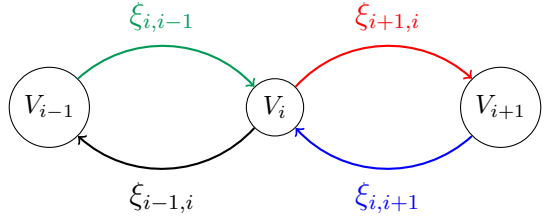
\begin{figure}
 \begin{center}
\begin{tikzpicture}
\node (One) at (-3,0) [shape=circle,draw] {$V_{i-1}$};
\node (Two) at (0,0) [shape=circle,draw] {$V_i$};
\node (Three) at (3,0) [shape=circle,draw] {$V_{i+1}$};
\def\myshift#1{\raisebox{-2.5ex}}
\draw [->,thick,color=black!100] (Two) to [bend left=45]  (One);
\def\myshift#1{\raisebox{1ex}}
\draw [->,thick,color=ForestGreen!100]      (One) to [bend left=45] (Two);
\def\myshift#1{\raisebox{1ex}}
\draw [->,thick,color=red!100]      (Two) to [bend left=45] (Three);
\draw [->,thick,color=blue!100]      (Three) to [bend left=45] (Two);
        \node[anchor=center,font=\large,ForestGreen!100] at (-1.5,1.2) {$\xi_{i,i-1}$};
        \node[anchor=center,font=\large,black!100] at (-1.5,-1.2) {$\xi_{i-1,i}$};
        \node[anchor=center,font=\large,red!100] at (1.5,1.2) {$\xi_{i+1,i}$};
        \node[anchor=center,font=\large,blue!100] at (1.5,-1.2) {$\xi_{i,i+1}$};
\end{tikzpicture}
\caption{Diagram corresponding to the asymmetric case.
It describes the simplest example defined by Eq.~\eref{exemple} and Eq.~\eref{exemple1}. The link
variables are four different uncorrelated random variables
(represented by different colors). The rules are the following, the
arrows exiting site $i$ correspond to the quantity which is
distributed to the neighbors. And the arrows coming in $i$ represent
the quantity that is given from the neighbors.}
               \label{asym}
        \end{center}
\end{figure}

Now let us consider the model with asymmetric links 
\begin{equation}
\xi_{ij}(t)\neq\xi_{ji}(t).
\end{equation} 
The model can be seen as acting on a directed graph where the
 incoming and outgoing links are two different independent random
 variables. As we saw in the previous section, exact conservation enforces
\begin{equation}\label{eq2}
V_i(t)=1-\sum_{k}\xi_{ki}(t){\mathrm K}_{k-i} \ \ \ \mathrm{then} \ \
\ \langle V\rangle=1-\langle \xi\rangle \|{\mathrm K}\|_{1}.
\end{equation}
 Let us write down an explicit example (ignoring the
 additive noise) for a line with three sites with closed boundary
 conditions and nearest-neighbors interactions. The evolution
 equation Eq.~\eref{eq1}, written in the matrix form Eq.~\eref{matrixform}, reads
  \begin{small}
 \begin{eqnarray}\label{exemple}
 \left(\begin{array}{c}u_{i-1}(t+1) \\u_i(t+1)
 \\u_{i+1}(t+1)\end{array}\right)= \left(\begin{array}{ccc}V_{i-1} &
 \xi_{i-1,i} & 0 \\  \textcolor{black}{\xi_{i,i-1}} & V_i &
 \textcolor{black}{\xi_{i,i+1}} \\ 0& \textcolor{black}{\xi_{i+1,i}}
 & V_{i+1}\end{array}\right)\left(\begin{array}{c}u_{i-1}(t)
 \\u_i(t) \\u_{i+1}(t)\end{array}\right),
  \end{eqnarray}
  \end{small}
where the column-stochastic condition is
 \begin{eqnarray}\label{exemple1}
&& V_{i-1}(t)=1-\textcolor{black}{\xi_{i,i-1}(t)},\\
&& V_i(t))=1-\textcolor{black}{\xi_{i-1,i}}-\textcolor{black}{\xi_{i+1,i}(t)},\nonumber\\
&& V_{i+1}(t)=1-\textcolor{black}{\xi_{i,i+1}}\nonumber.
 \end{eqnarray}
 We can check that $u_{i-1}(t+1)+u_{i+1}(t+1)+u_{i}(t+1)=u_{i-1}(t)+u_{i}(t)+u_{i+1}(t)$ \footnote{Let us remark that $\langle V\rangle=1-\langle \xi\rangle \|{\mathrm K}\|_{1}$ is not satisfied here because our example is not invariant by translation.}. In that case, the matrix ${\mathrm M}$ is not correlated, all the entries are independent.  It can be useful to write Eq.~ \eref{exemple} and Eq.~\eref{exemple1} as a diagram (see Fig~.(\ref{asym}) ). The rules are the following, the arrows exiting site $i$ correspond to the quantity which is distributed to the neighbors. And the arrows coming in $i$ represent the quantity that is given from the neighbors.
With exact conservation law, the variances of the
link and on-site noises are related by $\sigma_{V}^2=\|{\mathrm
K}\|^{2}\sigma_{\xi}^2$, where $\|{\mathrm K}\|=\sqrt{\sum_i
{\mathrm K}_{i}^2}$. Consequently, the correlations between the link
variables and the on-site variables take the form
\begin{eqnarray}\label{xi-V-asym}
\langle \xi_{rp}(t)V_0(t)\rangle-\langle \xi\rangle\langle V\rangle=-\sigma_{\xi}^2{\mathrm K}_r\delta_{p,0}.
\end{eqnarray}
The exact conservation law spatially correlates the link and on-site
 variables together. 
 Because of Eq.~\eref{eq2}, there is no cross-correlation between onsite
 variables at different sites
\begin{eqnarray}\label{asymetric}
 \langle V_r(t)V_0(t) \rangle-\langle V \rangle^2= \sigma_V^2\|{\mathrm K}\|^2 \delta_{r,0}.
 \end{eqnarray}
That shows that the onsite variables are simply uncorrelated in space. 
This asymmetric link assumption will be fully detailed in the next
sections, but it will be useful to explore the other cases 
for a better understanding of the differences. The asymmetric
case is the only one which does not create correlations inside the
matrix $M$ , as we shall see next.

\subsubsection{Symmetric links }

Now let us consider the same model but with symmetric links 
\begin{equation}
\xi_{ij}(t)=\xi_{ji}(t),
\end{equation}
and where one still have the deterministic relation Eq.~\eref{xiV} between the variables. The example of the past section can be written as well and the diagram becomes the following.
 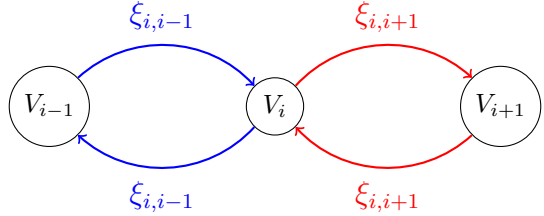
\begin{figure}
  \begin{center}
\begin{tikzpicture}
\node (One) at (-3,0) [shape=circle,draw] {$V_{i-1}$};
\node (Two) at (0,0) [shape=circle,draw] {$V_i$};
\node (Three) at (3,0) [shape=circle,draw] {$V_{i+1}$};
\def\myshift#1{\raisebox{-2.5ex}}
\draw [->,thick,color=blue!100] (Two) to [bend left=45]  (One);
\def\myshift#1{\raisebox{1ex}}
\draw [->,thick,color=blue!100]      (One) to [bend left=45] (Two);
\def\myshift#1{\raisebox{1ex}}
\draw [->,thick,color=red!100]      (Two) to [bend left=45] (Three);
\draw [->,thick,color=red!100]      (Three) to [bend left=45] (Two);
        \node[anchor=center,font=\large,blue!100] at (-1.5,1.2) {$\xi_{i,i-1}$};
        \node[anchor=center,font=\large,blue!100] at (-1.5,-1.2) {$\xi_{i,i-1}$};
        \node[anchor=center,font=\large,red!100] at (1.5,1.2) {$\xi_{i,i+1}$};
        \node[anchor=center,font=\large,red!100] at (1.5,-1.2) {$\xi_{i,i+1}$};
\end{tikzpicture}
\caption{Diagram corresponding to the symmetric case for the example Eq.~\eref{exemple}. The link variables are now the same $\xi_{ij}(t)=\xi_{ji}(t)$ between two neighbors. }
\label{symm}
 \end{center}
\end{figure}
In that situation, the matrix ${\mathrm M}$ is now doubly-stochastic because of the symmetry, but not uncorrelated anymore. The corresponding diagram can be seen on (see Fig~.\ref{symm} ) Indeed the symmetry induced some counter-diagonal correlations in the matrix ${\mathrm M}$. Here we still have the relation $\sigma_{V}^2=\|{\mathrm K}\|^{2} \sigma_{\xi}^2$
and
\begin{eqnarray}
\langle \xi_{rp}(t)V_0(t)\rangle-\langle \xi\rangle\langle V\rangle=\sigma_{\xi}^2{\mathrm K}_r\delta_{p,0},
\end{eqnarray}
like in the previous model. The main difference comes from the fact that
the $V$'s are now spatially correlated, and using Eq.~\eref{xiV} one finds
\begin{equation}\label{sym}
\langle V_r(t) V_0(t)\rangle-\langle V\rangle^2=\sigma_{\xi}^2\left(\|{\mathrm K}\|^2\delta_{r,0}+{\mathrm K}_r^2\right).
\end{equation}
We can observe, that the onsite variables are not delta-correlated anymore but are short-range correlated with a correlation kernel ${\mathrm K}_r^2$. The symmetry between links induces correlations between the onsite
variables at different positions.
 \footnote{If now we impose average conservation and ${\mathrm M_{ij}}$'s are Gaussian distributed, then
${\mathrm M}$ belongs to the Gaussian Orthonormal Ensemble
($\mathrm{GOE}$) \cite{mehta2004random}.}

\subsubsection{Isotropic links}

\begin{figure}[h!]
  \begin{center}
\begin{tikzpicture}
\node (One) at (-3,0) [shape=circle,draw] {$V_{i-1}$};
\node (Two) at (0,0) [shape=circle,draw] {$V_i$};
\node (Three) at (3,0) [shape=circle,draw] {$V_{i+1}$};
\def\myshift#1{\raisebox{-2.5ex}}
\draw [->,thick,color=red!100] (Two) to [bend left=45]  (One);
\def\myshift#1{\raisebox{1ex}}
\draw [->,thick,color=ForestGreen]      (One) to [bend left=45] (Two);
\def\myshift#1{\raisebox{1ex}}
\draw [->,thick,color=red!100]      (Two) to [bend left=45] (Three);
\draw [->,thick,color=blue!100]      (Three) to [bend left=45] (Two);
        \node[anchor=center,font=\large,ForestGreen] at (-1.5,1.2) {$J_{i-1}$};
        \node[anchor=center,font=\large,red!100] at (-1.5,-1.2) {$J_{i}$};
        \node[anchor=center,font=\large,red!100] at (1.5,1.2) {$J_{i}$};
        \node[anchor=center,font=\large,blue!100] at (1.5,-1.2) {$J_{i+1}$};
\end{tikzpicture}
\caption{Diagram corresponding to the isotropic case $\xi_{ij}(t)=J_j(t)$ for the example Eq.~\eref{exemple}. The outgoing links are equal (red) and the incoming are different uncorrelated random variables (blue and green). }
 \label{isot}
 \end{center}
\end{figure}
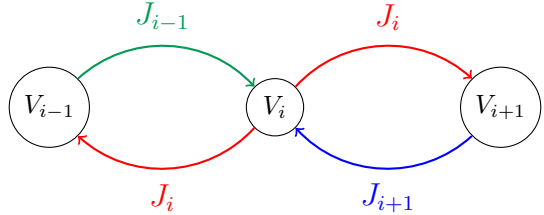

Let us consider the case where, the link variable does not depend on the first index
\begin{equation}
\xi_{ij}(t):=J_j(t),
\end{equation}
 of variance $\sigma_J^2$, then all the out-going links have
the same value, but the incoming links are independent. The diagram of Eq.~\eref{eq1}
in that case is Fig~.\ref{isot}. In that case, one has
\begin{equation}\label{eq4}
V_i(t)=1-J_{i}(t)\sum_{k}{\mathrm K}_{k-i}=1- J_i \|{\mathrm K}\|_{1}.
\end{equation}
and $\sigma_V^2=\sigma_{J}^2 \left(\|{\mathrm K}\|_{1}\right)^2 $
where $\|{\mathrm K}\|_{1}=\sum_i {\mathrm K}_i$. The exact conservation condition implies row-stochasticity of the
matrix ${\mathrm M}$.
In that case, the correlations between the link variables and the on-site variables are
\begin{eqnarray}
\langle J_{r}(t)V_0(t)\rangle-\langle J\rangle\langle V\rangle=-\sigma_{J}^2\|{\mathrm K}\|_{1}\delta_{r,0}.
\end{eqnarray}
 Because of Eq.~\eref{eq4}, there is no correlations between $V_r(t)$ and $V_0(t)$ at different sites
\begin{eqnarray}\label{isotrop}
 \langle V_r(t)V_0(t) \rangle-\langle V \rangle^2=\sigma_J^2\left(\|{\mathrm K}\|_{1}\right)^2\delta_{r,0}.
 \end{eqnarray}
Contrary to previous symmetric case, the isotropic assumption does not couple the on-site variables. This case is actually very interesting because of the form of the matrix ${\mathrm M}$, although it will not be discussed in this article.

\subsubsection{Gain noise}

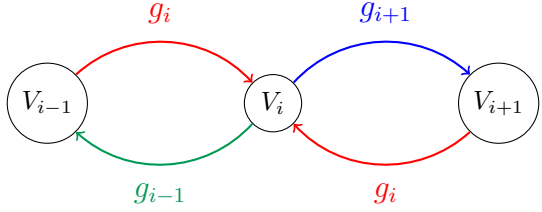
\begin{figure}[h!]
  \begin{center}
\begin{tikzpicture}
\node (One) at (-3,0) [shape=circle,draw] {$V_{i-1}$};
\node (Two) at (0,0) [shape=circle,draw] {$V_i$};
\node (Three) at (3,0) [shape=circle,draw] {$V_{i+1}$};
\def\myshift#1{\raisebox{-2.5ex}}
\draw [->,thick,color=ForestGreen] (Two) to [bend left=45]  (One);
\def\myshift#1{\raisebox{1ex}}
\draw [->,thick,color=red!100]      (One) to [bend left=45] (Two);
\def\myshift#1{\raisebox{1ex}}
\draw [->,thick,color=blue!100]      (Two) to [bend left=45] (Three);
\draw [->,thick,color=red!100]      (Three) to [bend left=45] (Two);
        \node[anchor=center,font=\large,red!100] at (-1.5,1.2) {$g_{i}$};
        \node[anchor=center,font=\large,ForestGreen] at (-1.5,-1.2) {$g_{i-1}$};
        \node[anchor=center,font=\large,blue!100] at (1.5,1.2) {$g_{i+1}$};
        \node[anchor=center,font=\large,red!100] at (1.5,-1.2) {$g_{i}$};
\end{tikzpicture}
\caption{Diagram corresponding to the gain noise case $\xi_{ij}(t)=g_i(t)$ for the example Eq.~\eref{exemple}. The incoming links are equal (red) and the outgoing are different independent random variables (blue and green). }
  \label{gainnoise}
 \end{center}
\end{figure}

The last case that one might be interested in here is  
\begin{equation}
\xi_{ij}(t):=g_i(t),
\end{equation}
of variance $\sigma_g^2$. This assumption can be seen as the opposite of the previous isotropic situation. Here we have
\begin{equation}
V_i(t)=1-\sum_{k}g_{k}(t){\mathrm K}_{k-i}.
\end{equation}
The diagram is now Fig~.\ref{gainnoise} (notice that the diagram is reversed compared to the previous isotropic case).
The relation between the variances is now $\sigma_{V}^2=\|{\mathrm
K}\|^{2}\sigma_{g}^2$. Here the exact conservation condition
implies column-stochasticity of the matrix ${\mathrm M}$.
The correlations between link variables and on-site variables take the form
\begin{eqnarray}
\langle g_{r}(t)V_0(t)\rangle-\langle g\rangle\langle V\rangle=-\sigma_{g}^2{\mathrm K}_r.
\end{eqnarray}
The correlations between the onsite variables are also different, one can show that
\begin{eqnarray}\label{gain}
\langle V_r(t) V_0(t)\rangle-\langle V\rangle^2=\sigma^2_{g}\|{\mathrm K}\|_1{\mathrm K}_r.
\end{eqnarray}
In this case, the $V's$ are spatially correlated by the kernel ${\mathrm K}_r$ du to the form
of the links and are also correlated with the links with the same kernel. In a sense, this specific form is closer to the symmetric
case, but the expressions of the correlations are different.

\subsection{Average conservation law}

The other scenario is to impose conservation in average at the thermodynamic limit $\langle
 m(t+1)\rangle=\langle m(t)\rangle$. This condition is a less restrictive constraint on the time-evolution of the system. This condition only imposes a relation in average between the onsite and link variables
 \begin{equation}
  \langle V\rangle=1-\langle\xi\rangle \|{\mathrm K}\|_{1},
 \end{equation}
where $\|{\mathrm K}\|_{1}=\sum_i {\mathrm K}(i)$. 
The diagonal and off-diagonal noises are now independent random quantities while there were linearly dependent when exact conservation was enforced. We are not repeating the analysis of correlations between noises in every cases here, let us just focus on the asymmetric model that one will be interested in in the rest of the article. Of course, when only average conservation is enforced, no relation between the variances $\sigma_{V}^2$ and $\sigma_{\xi}^2$ exists, and there is no cross-correlation between onsite and link variables
\begin{eqnarray}
\langle \xi_{rp}(t)V_0(t)\rangle-\langle \xi\rangle\langle V\rangle=0.
\end{eqnarray}
 Furthermore, the onsite variables $V$ are simply delta-correlated
\begin{eqnarray}\label{asymmetric}
 \langle V_r(t)V_0(t) \rangle-\langle V \rangle^2= \sigma_V^2 \delta_{r,0}.
 \end{eqnarray}
In the formulation Eq.~\eref{matrixform}, the
matrix ${\mathrm M}$ is not stochastic anymore $\sum_{i}{\mathrm
M}_{ij}\neq1$ but is average-stochastic $\sum_{i}\langle{\mathrm
M}_{ij}\rangle=1$. We shall see that those two different conservation conditions (exact or average) lead to
different results for the behavior of the coherence, that can be understood in terms of their
corresponding continuum space-time evolution equation.

\section{Threshold behavior of the network coherence}

\subsection{Phase diagram of the exactly conserved model}

As explained earlier, the behavior of the coherence of the system can be analyzed by computing the 2-point correlation function ${\mathrm G}_{r}(t)=\langle u_r(t)u_0(t)\rangle$ between $u_r(t)$ and $u_0(t)$ at same time $t$.  The calculation is shown here for the asymmetric
case with exact conservation law, but the steps are the same for all the different particular cases. In the asymmetric case
with exact conservation, we have Eq.~\eref{xi-V-asym} and  Eq.~\eref{asymetric}, therefore, it is straitforward to write down the time-evolution of the correlator ${\mathrm G}_{r}(t)$  as
\begin{eqnarray}\label{correl}
{\mathrm G}_{r}(t+1)&=&\langle \xi\rangle^2\sum_{k,l}{\mathrm K}_{r-k}{\mathrm K}_l{\mathrm G}_{k-l}(t) +2\langle V\rangle\langle \xi\rangle\sum_{k}{\mathrm K}_{r-k}{\mathrm G}_k(t)\nonumber\\
&+&\langle V\rangle^2  {\mathrm G}_{r}(t)+\delta_{r,0}{\mathrm G}_{0}(t)\left( \sigma_{\xi}^2\|{\mathrm K}\|^{2}
+ \sigma^2_{V}-\textcolor{black}{2\sigma_{\xi}^2{\mathrm K}^2_r}\right)\nonumber\\
&+& \sigma_n^2\delta_{r,0}.
\end{eqnarray}
From now on, we will focus on the analysis of this evolution equation for a short-range kernel but let us mention that this equation can be easily solved on the complete graph where the kernel takes the form  ${\mathrm K}_r=1-\delta_{r,0}$, the details of this
calculation will be presented elsewhere.

\subsubsection{Noise threshold in any dimension}

One could notice that Eq.~\eref{correl} can be written down as a convolution, then in Fourier space the result is a simple product plus a term that contains ${\mathrm G}_r(t)$ in $r=0$
\begin{eqnarray}\label{eq5}
\widehat{{\mathrm G}}({\bf q},t+1)&=&\left(\langle \xi\rangle \widehat{\mathrm K}({\bf q})+\langle V\rangle\right)^2\widehat{{\mathrm G}}({\bf q},t)
+{\mathrm G}_0(t)\left( \sigma_{\xi}^2 \|{\mathrm K}\|^{2} 
+ \sigma^2_{V}-\textcolor{black}{2\sigma_{\xi}^2\widehat{{\mathrm K}^2_r}}\right)
+\sigma_n^2,
\end{eqnarray}
with the following definition of the Fourier transform $\hat{u}_{\bf
q}(t)=\sum_{{\bf r}_i\in\mathbb{Z}^d}u_{\bf r_i}(t)e^{i {\bf q}{\bf
r}_i}$. As we saw earlier, the exact conservation law implies $\sigma_{V}^2=\sigma_{\xi}^2\|{\mathrm K}\|^{2}$ then Eq.~(\ref{eq5})
reads
\begin{eqnarray}\label{eqCorr}
\widehat{{\mathrm G}}({\bf q}, t+1)&=&\lambda({\bf q})^2\widehat{{\mathrm G}}({\bf q},t)+ 2{\mathrm G}_{0}(t)\sigma_{\xi}^2 \left(\|{\mathrm K}\|^{2}-\textcolor{black}{\widehat{{\mathrm K}^2_r}}\right)+\sigma_n^2\nonumber\\
&=&\lambda({\bf q})^2\widehat{{\mathrm G}}({\bf q},t)+ 2{\mathrm G}_{0}(t)\sigma_{\xi}^2 (\widehat{{\mathrm K}^2}(0)-
\widehat{{\mathrm K}^2}({\bf q}))+\sigma_n^2,
\end{eqnarray}
where we have
\begin{eqnarray}
\lambda({\bf q})&=&\langle \xi\rangle \widehat{\mathrm K}({\bf q})+\langle V\rangle\nonumber\\
&=&1-\langle \xi\rangle \left(  \|{\mathrm K}\|_{1}-\widehat{\mathrm K}({\bf q}) \right)\nonumber\\
&=&1-\langle \xi\rangle   \left(\widehat{\mathrm K}(0)-\widehat{\mathrm K}({\bf q})\right).
\end{eqnarray}
Now let us consider the continuum space limit of this problem, $u_{{\bf r}\in\mathbb{Z}^d}(t)\rightarrow u({\bf r}\in\mathbb{R}^d,t)$.
Let us notice that the case $\langle \xi\rangle=0$ is therefore rather particular, indeed it implies $\langle V\rangle=1\rightarrow\lambda({\bf q})=1$, and the evolution equation is trivial to solve \footnote{For symmetric links, this case is not trivial anymore.}, hence the correlator is always exponential for any value of the variances.
For  $\langle \xi\rangle\neq0$, the stationary solution $\widehat{{\mathrm G}}_{\mathrm st}({\bf q})$ is given by solving $\widehat{{\mathrm G}}({\bf q},t+1)-\widehat{{\mathrm G}}({\bf q},t)=0$, it follows
\begin{eqnarray}
\widehat{{\mathrm G}}_{\mathrm st}({\bf q})=\frac{2{\mathrm G}_{\mathrm st}(0) \sigma_{\xi}^2 \left(\widehat{{\mathrm K}^2}(0)-
\textcolor{black}{\widehat{{\mathrm K}^2}({\bf q})}\right)+\sigma_n^2}{1-\lambda({\bf q})^2},
\end{eqnarray}
where we used the notation ${\mathrm G}_{\mathrm st}(0):={\mathrm G}_{\mathrm st}({\bf r}=0)$. The self-consistent equation for ${\mathrm G}_{\mathrm st}(0)$ gives
${\mathrm G}_{\mathrm st}(0)=\frac{c_d\sigma_n^2 }{1-2\sigma_{\xi}^2 f_d}$
where, in the continuum-lattice limit, we have
\begin{eqnarray}\label{c}
c_d=\frac{1}{(2\pi)^d}\int_{-\infty}^{\infty}\frac{\mathrm{d}^{d}{\bf q}}{1-\lambda({\bf q})^2 } ,
\end{eqnarray}
and
\begin{eqnarray}\label{f}
 f_d=\frac{1}{(2\pi)^d}\int_{-\infty}^{\infty} \mathrm{d}^{d}{\bf q} \ \frac{\widehat{{\mathrm K}^2}(0)-
\widehat{{\mathrm K}^2}({\bf q})}{1-\lambda({\bf q})^2 }.
\end{eqnarray}
Until now, the calculation holds for any integrable kernel, and we
shall now focus on a local kernel  which scales as a power-law in
Fourier space $\widehat{\mathrm K}({\bf q})\sim{\bf q}^{2\theta}$,
the simplest being the Laplacian corresponding to $\theta=1$. For a local kernel, one has
$\lambda({\bf q})^2\sim 1+p\langle\xi\rangle{\bf q}^{2\theta}$ where $p$ is a constant.  The
integral $c_d$ converges in $d>2\theta$ and $f_d$ is convergent in any
dimension. The integrals $c_d$ and $f_d$ can be defined, by dimensional
regularization, for any real $d$ \footnote{then the calculation also holds
for a fractal graph of non-integer dimension $d$}. The stationary
solution exists if and only if $1-2f_d \sigma_{\xi}^2 >0$. The
critical value 
\begin{equation}
\sigma^2_\mathrm{c}=\frac{1}{2f_d},
\end{equation}
 where $f_d$ is given by Eq.~\eref{f}, is the maximum value of the
variance $\sigma^2_\xi$ such that the stationary state is reachable and such that the system is stable. The constant
$f_d$ depends only on the explicit form of the kernel, the mean of the
link variables $\langle \xi\rangle$ and the dimension $d$ of the space.
The value of this threshold can be tuned by changing the value of
$\langle \xi\rangle$. So a first result shows that, in that situation, the stability margin Eq.~\eref{stabilitymargin} is always positive in any dimension, therefore the system can support some amount of multiplicative noise and reach its stationary state. The full explicit form of the stationary correlator can be computed. For $\sigma_{\xi}^2<\sigma_\mathrm{c}^2$ and for $d>2\theta$ the stationary solution can be written
\begin{eqnarray}\label{eq6}
{\mathrm G}_{\mathrm st}({\bf r})=\frac{2\sigma_{\xi}^2}{(2\pi)^d}\frac{c_d\sigma_n^2 }{1-2\sigma_{\xi}^2 f_d}\int\mathrm{d}^{d}{\bf q} \frac{\widehat{{\mathrm K}^2}(0)-
\textcolor{black}{\widehat{{\mathrm K}^2}({\bf q})} }{1-\lambda({\bf q})^2}e^{-i{\bf q}{\bf r}  }.
\end{eqnarray}
The stationary solution converges to a finite value in $d>2\theta$. As usual, the divergence is logarithmic at $d_\mathrm{c}$.
The behavior can be easily understood in a renormalization group (RG) language. In the phase $\sigma_{\xi}^2<\sigma_\mathrm{c}^2$, we can define the quantity $ A^{{\rm ex}}_{\sigma^2}$
\begin{eqnarray}
A^{{\rm ex}}_{\sigma^2}= \frac{1}{1-2\sigma^2_\xi f_d}.
\end{eqnarray}
This quantity, which goes to $1$ when $\sigma_\xi^2\rightarrow0$ and
goes to $\infty$ when $\sigma_\xi^2\rightarrow\sigma^2_\mathrm{c}$, can be
seen as a coefficient which renormalizes the additive noise. Indeed we have ${\mathrm G}_{\mathrm st}(0)= A^{{\rm ex}}_{\sigma^2}\sigma_n^2f_d$, which is, up to the constant $A^{{\rm ex}}_{\sigma^2}$, the stationary solution of a pure additive system ($\sigma_\xi^2=0$), as we shall see in the next paragraph. What it
means, is that below the noise threshold, the
randomness of the link and onsite variables are irrelevant in a RG
sense and the system renormalizes to
a pure additive model
 with a renormalized additive noise $\tilde n_i(t)=\sqrt{A^{{\rm ex}}_{\sigma^2}}n_i(t)$ where $n_i(t)$ is the original additive noise. The correlator equation becomes
\begin{eqnarray}\label{weakcoupling}
\widehat{{\mathrm G}}({\bf q},t+1)=\lambda({\bf q})^2\widehat{{\mathrm G}}({\bf q},t)+A^{{\rm ex}}_{\sigma^2}\sigma_n^2.
\end{eqnarray}
Since $\sigma_\xi^2$ is the only relevant parameter dictating the large-scale behavior of the system, the effective description Eq.~(\ref{weakcoupling}) of the system below $\sigma^2_\mathrm{c}$  is still valid in the non-stationary regime $d\leqslant2\theta$.  The
full space-time-dependent solution of Eq.~(\ref{weakcoupling}), for a
kernel of the form $\widehat{\mathrm K}({\bf q})\sim{\bf
q}^{2\theta}$ verifies the following scaling form
\begin{equation}\label{eq8}
{\mathrm G}({\bf r},t)=A^{{\rm ex}}_{\sigma^2}\sigma_n^2{\bf r}^{2\theta-d}\Psi \left(\frac{t}{{\bf r}^{2\theta}}\right).
\end{equation}
The form of the correlator is valid for any dimension below the threshold.
Here $\Psi(y)$ is a scaling function with properties that
$\Psi(y)\rightarrow const$ as $y\rightarrow\infty$ and
$\Psi(y)\rightarrow y^{(2\theta-d)/2\theta}$ as $y\rightarrow 0$.
This scaling form Eq.~(\ref{eq8}) is the so-called Family-Viczek
($\mathrm{FV}$) scaling  \cite{family1985scaling}, well-known in
statistical physics of interfaces. The case $\theta=1$ and
$\theta=2$ are respectively the Edwards-Wilkinson (EW) and
Mullins-Herring (MH) universality classes
\cite{barabasi1995fractal}. The upper-critical dimension of this
system is $d_\mathrm{c}=2\theta$, above that dimension, the correlator
converges to a finite value. 
For $\sigma_{\xi}^2<\sigma_\mathrm{c}^2$ and for $d\leqslant2\theta$ the
behavior is a power-law following Eq.~(\ref{eq8}). The correlator
grows as $t^{(2\theta-d)/2\theta}$ and the stationary solution is
never reached for an infinite system. This exponent is named $2\beta$ in the context of
growing interfaces and is equal to $\beta=1/4$ (resp $\beta=3/8$)
for the EW class (resp. MH). The exponent increases with $\theta$.

\subsubsection{Finite size dependence of the coherence}

Now we can extend the result on the large-time coherence scaling for the Laplacian algorithm ($\theta=1$) sketched in the introduction.
From the result on the calculation of ${\mathrm G}({\bf r},t)$ below the noise threshold, one can translate the informations in terms of the time-dependance of $C_N(t)$ and the stationary value $C_N^{\infty}$ of the network coherence defined (see Eq.~\eref{coherence} and Eq.~\eref{timecoherence}) in the introduction for a system of $N$ agents. The finite-size behavior of the coherence can be extracted by introducing a cut-off in Fourier space in Eq.~\eref{c} and Eq.~\eref{f}. For a system of size $L$ with $N=L^d$ nodes, the coherence is behaving for short times ($t\ll  L^{2}$)
\begin{equation}
C_N(t)\propto\frac{1}{N}\int{\mathrm d}^d{\bf r}{\mathrm G}({\bf r},t)\sim t^{(2-d)/2}.
\end{equation}
So for short times, the coherence grows as a power-law, independently of system-size. The coherence reaches then a stationary value which scales with system-size as
\begin{equation}
C^{\infty}_N\propto\frac{1}{N}\int{\mathrm d}^d{\bf r}{\mathrm G}_{\mathrm st}({\bf r}) \sim N^{\frac{2-d}{d}},
\end{equation}
for any $d$. This value is reached within a time-scale $t_c\sim L^{2}=N^{2/d}$. The value of the infinite coherence $C^{\infty}_N$ grows unboundedly in $d<3$ and converges to a finite value in $d>2$, in agreement with Eq.~\eref{scalingC}. In terms of performance, in $d>2$ the algorithm is still capable to perform its averaging task even with the presence of multiplicative noise, as long as the variance is below the threshold $\sigma^2_\mathrm{c}$. 

\subsubsection{Exponential regime above the noise threshold }

Now that we have understood the behavior of the correlator below the
threshold, where the large-scale behavior was dictated by the additive noise,
one can study the behavior in the other regime, where the behavior
will be controlled by the parameter $\sigma_\xi^2$. For
$\sigma_{\xi}^2\geqslant\sigma^2_{{\rm c}}$, the additive noise is
irrelevant since we expect an exponential growth of the correlator,
and the equation Eq.~\eref{eqCorr} can be written as  
\begin{eqnarray}\label{EV}
\widehat{{\mathrm G}}({\bf q},t+1)={\mathrm D}(\widehat{{\mathrm G}}({\bf q},t)).
\end{eqnarray}
The large-time behavior of this equation is dominated by the largest
eigenvalue $\lambda_{\mathrm{max}}$ of the operator ${\mathrm D}$. If
$\lambda_{\mathrm{max}}=1$, the
$\mathrm{FV}$ scaling Eq.~\ref{eq8} holds. If
$\lambda_{\mathrm{max}}>1$ then the correlator behaves exponentially as
\begin{eqnarray}\label{exp}
\widehat{{\mathrm G}}({\bf q},t)\sim \lambda_{\mathrm{max}}^t \widehat{g}({\bf q}),
\end{eqnarray}
where $\widehat{g}({\bf q})$ has to determined self-consistently. Inserting Eq.~\eref{exp} inside Eq.~\eref{eq5}, one ends up with 
\begin{eqnarray}
\widehat{g}({\bf q})=\frac{2\sigma_{\xi}^2(\widehat{{\mathrm K}^2}(0)-
\widehat{{\mathrm K}^2}({\bf q})}{
\lambda_{\mathrm{max}}-\lambda({\bf q})^2}.
\end{eqnarray}
 Now let us write a condition on the form of the largest eigenvalue
$\lambda_{\mathrm{max}}$ of the operator ${\mathrm D}$. We have
\begin{eqnarray}
\widehat{{\mathrm G}}({\bf q},t)=2\sigma_{\xi}^2\frac{\widehat{{\mathrm K}^2}(0)-
\widehat{{\mathrm K}^2}({\bf q})
}{\lambda_{\mathrm{max}}-\lambda({\bf q})^2}{\mathrm G}(0,t)=\widehat{g}({\bf q}){\mathrm G}(0,t).
\end{eqnarray}
%
%
   \begin{figure}
    \begin{center}
  \begin{tikzpicture}
  \draw[->] (0,0) -- (6,0) node[anchor=north] {$\sigma_{\xi}^2$};
\draw	(0,0) node[anchor=north] {0}
		(2,0) node[anchor=north] {{\scriptsize $\sigma_\mathrm{c}^2=\frac{1}{2f_d}$}}
		(5.65,2.1) node[anchor=north] {{\scriptsize $d_\mathrm{c}=2$}};
\draw	(1.5,3) node{{\scriptsize Finite }}
						(1.5,2.6) node{{\scriptsize $C_{N}^{\infty}=\frac{c_d\sigma_n^2 }{1-2\sigma_{\xi}^2 f_d}$}}

		(1.55,1.5) node{{\scriptsize Algebraic and stationary }}
								(1.13,1) node{{\scriptsize $C_{N}(t\ll  t_c)\sim t^{2\beta}$}}
								(1.31,0.7) node{{\scriptsize $C_{N}(t \gg\ t_c)\sim N^{\frac{2-d}{d}}$}}

		(5,1) node{{\scriptsize Exponential}}
				(5,0.7) node{{\scriptsize $C_{N}(t)\sim \lambda_{\mathrm{max}}^t$}}
		(5.3,3) node{{\scriptsize Exponential}}
						(5.3,2.7) node{{\scriptsize $C_{N}(t)\sim \lambda_{\mathrm{max}}^t$}};
\draw[->] (0,0) -- (0,4) node[anchor=east] {$d$};
\draw[thick,dashed,blue] (0,1.7) -- (6,1.7) ;
\draw[thick,red] (2,0) .. controls (4.5,2) and (4.5,3)  .. (4.5,4);
  \end{tikzpicture}
\caption{Phase diagram in the exact conservation case on a
$d$-dimensional lattice for the Laplacian kernel $\theta=1$ with $N$ sites. On the left of the critical line (red line)
$\sigma_{\xi}^2<\sigma^2_{{\rm c}}$, the behavior of the coherence
follows the $\mathrm{FV}$ scaling Eq.~\eref{eq8}, {\it i.e.}
algebraic (with exponent $\beta=(2-d)/4$) for short time then stationary at large time, in low $d$ (below the blue dashed line) and finite for higher $d$. On the other side of the line, the coherence grows exponentially for any $d$. In that situation, the algorithm remains resilient to multiplicative noise below the noise threshold, showing that the stability margin is positive in any dimension.}
\label{tab1}
 \end{center}
\end{figure}
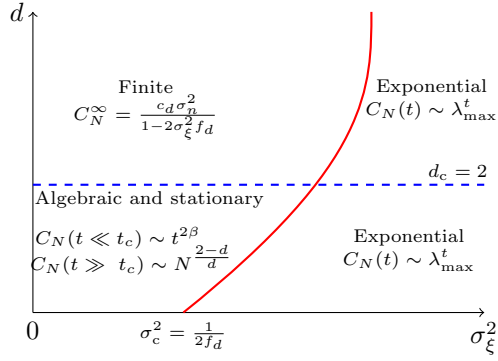
Now using $(2\pi)^{-d} \int \mathrm{d}^{d}{\bf q}\widehat{{\mathrm G}}({\bf q},t)={\mathrm G}(0,t)$ we end up with a condition on the largest eigenvalue $\lambda_{\mathrm{max}}$
\begin{eqnarray}\label{eq9}
\int \frac{\mathrm{d}^{d}{\bf q}}{(2\pi)^d}\widehat{g}({\bf q})=1,
\end{eqnarray}
where
\begin{equation}
\widehat{g}({\bf q})=\frac{\mathrm{d}^{d}{\bf q}}{(2\pi)^d} \ \frac{
\textcolor{black}{\widehat{{\mathrm K}^2}(0)-\widehat{{\mathrm K}^2}({\bf q})} }{\lambda_{\mathrm{max}}-\lambda({\bf q})^2},
\end{equation}
and  $\lambda({\bf q})^2\approx1+p \langle\xi\rangle {\bf
q}^{2}$.  The integral Eq.~\eref{eq9} when
$\lambda_{\mathrm{max}}\rightarrow\ 1$, converges in any $d$ by
dimensional regularization, meaning that the $\mathrm{FV}$ scaling holds in any $d$ below the
threshold. The behavior of the coherence $C(t)$ in the regime $\sigma_{\xi}^2\geqslant\sigma^2_{{\rm c}}$ is always exponential in any dimension and for any system-size. Those different regimes can be understood by looking at the
corresponding continuum equation describing the large-scale
fluctuations of  Eq.~\eref{eq1}. Without loss of generalities, let us focus on the Laplacian kernel from now on. We showed previously that, when exact conservation is enforced, it exists a deterministic relation between the onsite and link noises Eq.~\eref{eq2}.
Therefore the asymptotic behavior of our model is governed by a continuum space-time equation with only one multiplicative source of randomness and additive noise
 \begin{equation}
\partial_{t}u({\mathbf x},t)=\partial_{{\mathbf x}}\left[\xi({\mathbf x},t)\partial_{{\mathbf x}}u({\mathbf x},t)\right]+n({\mathbf x},t).
\end{equation}
From our analysis, one shows that there are physically two different regimes. The first regime, is when $\xi({\mathbf x},t)$ is irrelevant at large distance and one ends up with the large-scale behavior of the Edwards-Wilkinson equation
 \begin{equation}
\partial_{t}u({\mathbf x},t)=\partial^2_{{\mathbf x}}u({\mathbf x},t)+\tilde n({\mathbf x},t),
\end{equation}
where the additive noise $\tilde n({\mathbf x},t)$ is renormalized by the value of the variance $\sigma^2_\xi$ of the multiplicative noise as  $\tilde n({\mathbf x},t)=\sqrt{A^{{\rm ex}}_{\sigma^2}}n({\mathbf x},t)$, where $n({\mathbf x},t)$ is the original additive noise. The other regime is exponential (see Eq.~\eref{exp}), where the large-scale behavior is dominated by the multiplicative noise $\xi({\mathbf x},t)$ and
where the additive noise does not change the asymptotic behavior. This two
regimes are separated by a threshold $\sigma_{{\rm c}}^2$ that is finite in any dimension and equal to $\sigma^2_\mathrm{c}=\frac{1}{2f_d}$ where $f_d$ is the integral given by Eq.~\eref{f}.

\subsection{Phase diagram of the average conserved model}

Now let us look at the our system when the time-evolution of $m(t)$ is conserved in average. We will see that the behavior changes drastically when one imposes average conservation, especially in high dimensions. In the aymmetric case with average conservation we have no correlation between the onsite and link variables, therefore
the evolution of the correlator  ${\mathrm G}_{r}(t)$ takes the following form
\begin{eqnarray}
{\mathrm G}_{r}(t+1)&=&\langle \xi\rangle^2\sum_{k,l}{\mathrm K}_{r-k}{\mathrm K}_l{\mathrm G}_{k-l}(t) +2\langle V\rangle\langle \xi\rangle\sum_{k}{\mathrm K}_{r-k}{\mathrm G}_k(t)\nonumber\\
&+&\langle V\rangle^2  {\mathrm G}_{r}(t)+\delta_{r,0}{\mathrm G}_{0}(t)\left( \sigma_{\xi}^2\|{\mathrm K}\|^{2}
+ \sigma^2_{V}\right)+ \sigma_n^2\delta_{r,0}.
\end{eqnarray}
Let us focus on the analysis of the evolution equation in the next section. The steps of calculation are very similar to the exactly conserved system, therefore one jumps immediately to the analysis of the solution.

\subsubsection{Noise threshold in high dimensions}

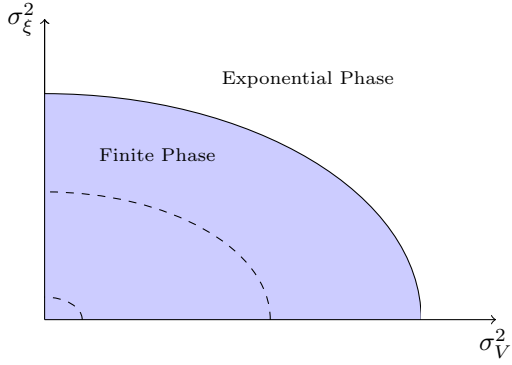
\begin{figure}
   \begin{center}
  \begin{tikzpicture}
  \draw[->] (0,0) -- (6,0) node[anchor=north] {$\sigma^2_{V}$};
  \draw	(1.5,1.2) node{{\scriptsize Finite Phase}};
    \draw	(3.5,3.2) node{{\scriptsize Exponential Phase}};		;;
\draw[->] (0,0) -- (0,4) node[anchor=east] {$\sigma_{\xi}^2$};
          \clip (0cm,0) rectangle (5,4cm);
    \draw[thick, black]  (0,0) ellipse(5cm and 3cm);
          \clip (0cm,0) rectangle (5,4cm);
          \fill[blue!20]  (0,0) ellipse(5cm and 3cm);
            \draw	(1.5,2.2) node{{\scriptsize Finite Phase}};
              \draw[->] (0,0) -- (6,0) node[anchor=north] {$\sigma^2_{V}$};
\draw[->] (0,0) -- (0,4) node[anchor=east] {$\sigma_{\xi}^2$};
              \draw[dashed, black]  (0,0) ellipse(3cm and 1.7cm);
                            \draw[dashed, black]  (0,0) ellipse(0.5cm and 0.3cm);
  \end{tikzpicture}
\caption{There is a critical line $f(\sigma^2_{V},\sigma^2_{\xi})$ in the plane $(\sigma^2_{V},\sigma^2_{\xi})$ which is parametrized by the ellipse equation $\frac{\sigma^2_{V}}{\sigma_{{\rm c}}^2}+\frac{\sigma_{\xi}^2}{\sigma_{{\rm c}}^2}\|{\mathrm K}\|^2=1$. In the white domain defined by $f(\sigma^2_{V},\sigma^2_{\xi})\geqslant1$, the algorithm becomes unstable and the coherence is exponential. In the blue domain defined by $f(\sigma^2_{V},\sigma^2_{\xi})<1$, it is finite and the algorithm reaches a global consensus. Let us mention that the critical line exists only in high dimensions $d>2$, the size of the ellipse decreases (the dashed ellipses) as $d\rightarrow d_c$ and eventually shrinks down to a single point at $d_c$.  The system is then completely unstable for any non-zero variances.
}
\label{ellipse}
 \end{center}
\end{figure}

Let us again focus on the Laplacian kernel without loss of generality. The self-consistent equation for ${\mathrm G}_{\mathrm st}(0)$ gives 
\begin{eqnarray}\label{stationary}
{\mathrm G}_{\mathrm st}(0)=\frac{c_d\sigma_n^2}{1-c_d \sigma^2},
\end{eqnarray}
where $c_d$ is given by the same integral Eq.~\eref{c} as in the previous section and  $\sigma^2=\sigma^2_{V}+\sigma_{\xi}^2 \|{\mathrm K}\|^2$. The stationary solution exists if and only if $1-c_d \sigma^2 >0$. The critical value of the variance being  
\begin{equation}
\sigma^2_{{\rm c}}=\frac{1}{c_d}.
\end{equation}
There is a critical line $f(\sigma^2_{V},\sigma^2_{\xi})$ in the plane $(\sigma^2_{V},\sigma^2_{\xi})$ which is parametrized by the ellipse (see Fig~.(\ref{ellipse})) equation 
\begin{equation}
\frac{\sigma^2_{V}}{\sigma_{{\rm c}}^2}+\frac{\sigma_{\xi}^2}{\sigma_{{\rm c}}^2}\|{\mathrm K}\|^2=1.
\end{equation}
 The constant $c_d$ depends only of the explicit form of the kernel, the mean of the link variables $\langle \xi\rangle$ and the dimension $d$ of the space. 
 Because of the divergence of $c_d$ in $d\leqslant2$, there is no threshold in low dimensions and the system is always controlled by the variables $\xi$'s and $V$'s as we will see later on. In $d\leqslant2$, the stability margin Eq.~\eref{stabilitymargin} becomes infinitesimal and any non-zero amount of multiplicative noise makes the algorithm unstable. 
The full space-dependance of the stationary solution below $\sigma_\mathrm c^2$ and in $d>2$ takes the following form
\begin{eqnarray}\label{eqq12}
{\mathrm G}_{\mathrm st}({\bf r})&=&\frac{\sigma_n^2}{(2\pi)^d}\left( \frac{1}{1-c_d\sigma^2} \right) \int_{-\infty}^{\infty} {\mathrm d}^d {\bf q} \frac{e^{- i {\bf q}{\bf r} }}{1-\lambda({\bf q})^2}.
\end{eqnarray}
The picture is almost the same as in the exactly conserved system where below the threshold, the system behaves as a pure additive model Eq.~\eref{weakcoupling}. In this regime, the renormalization factor becomes 
\begin{eqnarray}
A^{{\rm av}}_{\sigma^2}= \frac{1}{1-c_d\sigma^2}.
\end{eqnarray}
Below $\sigma^2_{{\rm c}}$, the system is also described by the pure additive model
Eq.~\eref{weakcoupling} with $\tilde n_i(t)=\sqrt{A^{{\rm
av}}_{\sigma^2}}n_i(t)$ and the $\mathrm{FV}$ scaling
Eq.~\eref{eq8}) holds with the same set of exponents. The main
difference is the $\mathrm{FV}$ scaling holds only in $d>2$,
where the exponent $\beta$ goes to zero, and where the stationary
solution Eq.~\eref{eqq12} converges to a finite value Eq.~\eref{stationary}. We have actually a critical line  (see Fig~.(\ref{ellipse})) in the plane $(\sigma^2_{V},\sigma^2_{\xi})$. The behavior is the same anywhere below the line where the large-scale dynamic is described by Eq.~\eref{weakcoupling}.  Because there is no relation between the variances $\sigma^2_{V}$ and $\sigma^2_{\xi}$, we can send independently $\sigma^2_{V}$ or $\sigma^2_{\xi}$ to zero. The large-scale dynamic above the threshold can then be described by
\begin{equation}\label{SHE}
u_i(t+1)=\langle\xi\rangle\sum_{j}{\mathrm K}_{i-j}u_j(t)+\tilde{V}_i(t)u_i(t),
\end{equation}
with $\tilde{V}_i(t)$ has a renormalized variance
$\sigma^2_{V}+\sigma_{\xi}^2 \|{\mathrm K}\|^2$. 
The behavior of our system when average conservation is imposed, varies greatly from the exactly conserved algorithm, where the system could not be described by Eq.~\eref{SHE} due to the relation between the variances $\sigma^2_{V}$ and $\sigma_{\xi}^2$. The eigenvalue equation for this model can also be written as Eq.~\eref{EV}. If $\lambda_{\mathrm{max}}>1$ then the correlator behaves as
$\widehat{{\mathrm G}}({\bf q},t)\sim \lambda_{\mathrm{max}}^t \widehat{g}({\bf q})$, where 
\begin{eqnarray}
\widehat{g}({\bf q})=\frac{\sigma^2_{V}+\sigma_{\xi}^2 \|{\mathrm K}\|^2}
{\lambda_{\mathrm{max}}-\lambda({\bf q})^2}.
\end{eqnarray}
  The condition on $\lambda_{\mathrm{max}}$ is now
\begin{eqnarray}\label{eq10}
\int \frac{\mathrm{d}^{d}{\bf q}}{(2\pi)^d}\widehat{g}({\bf q})=1.
\end{eqnarray}
The convergence properties of this integral are fairly different
 than the one found in the previous section. The integral Eq.~\eref{eq10} when
 $\lambda_{\mathrm{max}}\rightarrow\ 1$ converges in $d>2$.  Thus in
 $d\leqslant2$ there is no stationary state and the correlator grows
 exponentially. In $d>2$ there is a noise threshold $\sigma^2_{{\rm
 c}}$, below this value the behavior is algebraic, and above it is
 exponential for any $\sigma^2$.
In that case, the phase transition occurs at $d>2$, indeed in
$d\leqslant2$ the system remains exponential for any value of $\sigma^2$.  
  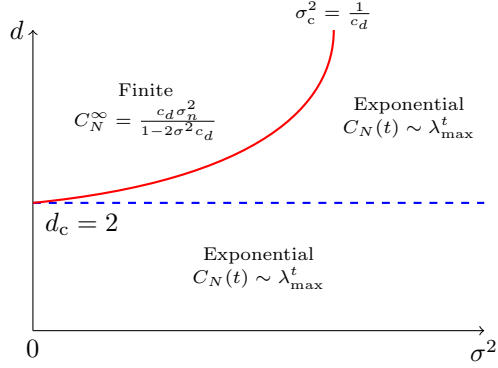
\begin{figure}
   \begin{center}
  \begin{tikzpicture}
  \draw[->] (0,0) -- (6,0) node[anchor=north] {$\sigma^2$};
\draw	(0,0) node[anchor=north] {0}
		(4,4.5) node[anchor=north] {{\scriptsize $\sigma_{{\rm
 c}}^2=\frac{1}{c_{d}}$}}
		(0.65,1.7) node[anchor=north] {$d_{{\rm
 c}}=2$};
\draw	(1.5,3.2) node{{\scriptsize Finite}}
						(1.5,2.8) node{{\scriptsize $C^{\infty}_N=\frac{c_d\sigma_n^2 }{1-2\sigma^2 c_d}$}}
		(3,1) node{{\scriptsize Exponential}}
								(3,0.7) node{{\scriptsize $C_N(t)\sim \lambda_{\mathrm{max}}^t$}}
		(5,3) node{{\scriptsize Exponential}}
								(5,2.7) node{{\scriptsize $C_N(t)\sim \lambda_{\mathrm{ max}}^t$}};;
\draw[->] (0,0) -- (0,4) node[anchor=east] {$d$};
\draw[thick,dashed,blue] (0,1.7) -- (6,1.7) ;
\draw[thick,red] (0,1.7) .. controls (3,2) and (4,3)  .. (4,4);
  \end{tikzpicture}
\caption{Phase diagram in the average conservation case for the Laplacian kernel $\theta=1$. Here the scenario is rather different than the exact conserved case in dimension $d$. In low $d$, the behavior of the coherence is always exponential for any system-size. In higher $d$ there is a transition between a finite phase and an exponential regime at the critical value  $\sigma_{{\rm
 c}}^2$.}
\label{tab2}
 \end{center}
\end{figure}
The low $d$ phase is the strong
coupling regime and in higher $d$, there is two phases depending on
the value of $\sigma^2$. Let us finally notice that, contrary to the exact conservation case, here the phase transition may happen at an infinitesimal value of $\sigma$, just above the critical dimension $d_{{\rm c}}$. The complete phase diagram is shown in Fig~.(\ref{tab2}).
Two regimes are present, just as before, the EW limit where
the variance $\sigma^2$ is smaller than the
critical value and where the multiplicative noise is irrelevant. This happens in
low dimensions. The other regime can be shown to be described by
the stochastic heat equation (SHE)
 \begin{equation}
\partial_{t}u({\mathbf x},t)=\partial^2_{{\mathbf x}}u({\mathbf x},t)+{\tilde V}({\mathbf x},t)u({\mathbf x},t).
\end{equation}
This equation is known to describe the partition function
$u({\mathbf x},t)$ of a directed polymer in a quenched random
potential ${\tilde V}({\mathbf x},t)$ and the height $h({\mathbf
x},t)\propto \log u({\mathbf x},t)$ of a random interface verifying
the Kardar-Parisi-Zhang equation (see \cite{krug1991kinetic} for details or
\cite{halpin1995kinetic} for even more details). 

\subsubsection{Robustness of the coherence and finite-size dependance}

The analysis of the correlator ${\mathrm G}({\mathbf r},t)$ tells us that in low dimensions, the consensus algorithm is extremely sensitive to multiplicative noise. The network coherence becomes essentially exponential as soon as a finite value of multiplicative noise is introduced in the system, making the algorithm unstable and unable to perform its task. The stability margin, is essentially zero in $d<3$. In higher dimensions $d>2$, the stability margin becomes is non-zero, and the system remains stable as long as the noise variance remains below the threshold. The higher the dimension and the bigger the stability margin is.  From that result on the calculation of ${\mathrm G}({\bf r},t)$ below the noise threshold, one can translate the information in terms of the stationary value $C_N^{\infty}$ of the network coherence defined (see Eq.~\eref{coherence}). The finite-size behavior of the coherence can be extracted by introducing a cut-off in Fourier space in Eq.~\eref{c}. For a system of size $L$ with $N=L^d$ nodes, the coherence is reaching a stationary value which scales as system-size as 
\begin{equation}
C^{\infty}_N\propto\frac{1}{N}\int{\mathrm d}^d{\bf r}{\mathrm G}_{\mathrm st}({\bf r}) \sim N^{\frac{2-d}{d}}.
\end{equation}
In that context, in $d>2$ the algorithm is still capable to perform its task, even with the presence of multiplicative noise, as long as the variance is below the threshold $\sigma^2_\mathrm{c}$. In lower dimensions, the stability margin is inexistent and there is no stationary state. An infinitesimal amount of multiplicative noise makes the system exponentially unstable, the system is thus infinitely fragile.

\subsection{A word on the other forms of symmetry}

 As we said earlier in the article, the asymmetric link case, is the simplest case where the link and onsite variables are uncorrelated in space, see Eq.~\eref{asymetric}), regardless the type of conservation condition that we consider. Therefore, this case leads to a rather simple equation for the time-evolution of the correlator, and the calculation can be done for any kernel. The other cases might be more or less involved depending on the form of the links and the conservation condition, indeed when one
enforces symmetry between link variables for example, the onsite random variables $V_i(t)$ become spatially correlated (see Eq.~\eref{sym}) leading to different asymptotic behaviors. Nevertheless, the method that we have proposed here can be applied directly to those situations, leading to the explicit form of the noise threshold. The details of those models will be detailed elsewhere although let us finally mention, that the crucial ingredient in the different universal behaviors that we have observed is the presence of some form of a conservation law, and we believe that the specific details of the system will not change the dimensional behaviors and phase diagrams Fig~.(\ref{tab1}) and Fig~.(\ref{tab2}), but could have some relevance for more specific applications to real consensus algorithms and communication networks.

\section{Summary and conclusions}

In this paper, we have quantified the performance of a consensus algorithm or a distributed system by studying the network coherence Eq.~\eref{coherence} of a system of $N$ agents
 \begin{eqnarray}
C_N(t)=\frac{1}{N}\int{\mathrm d}^d{\bf r}\left({\mathrm G}({\bf r},t)-\langle m(t) \rangle^2\right),
\end{eqnarray}
in the large-scale limit $N\rightarrow\infty$ and where ${\mathrm G}({\bf r},t)$ is the local 2-point correlation $\langle u({\bf r},t)u({\bf 0},t)\rangle$. We were interested in the behavior of this quantity for a system with diverse sources of uncorrelated multiplicative and additive noise (see Eq.~\eref{eq1}) in a lattice of dimension $d$ in the continuum-space limit. We showed how the quantity ${\mathrm G}({\bf r},t)$ behaves in both cases where one imposes conservation of $m(t)$ either exactly or in average. 
Let us summarize the results of the continuum-space calculation of
the time-behavior of the network coherence $C_N(t)$, for a system of $N$ nodes, when varying
the strength of the link and onsite random variables. The general behavior in the exact
conservation case, is that in any dimension, below a critical
value $\sigma_{{\rm
 c}}^2$, the network coherence grows algebraically then reaches
the stationary state at large time, and above $\sigma_{{\rm
 c}}^2$, the network coherence grows exponentially, see Fig~.(\ref{tab1}). The stability margin is always non-zero in any dimension, and the system can remain stable as long as the variance of the noise is below the threshold. The other case is when average
conservation of $m(t)$ is enforced. In that case, the link and onsite
variables are independent, making the phenomenology richer
than the exact conservation case. In low dimensions, the
network coherence grows exponentially, for any value of the variances
$\sigma^2_{V}$ and $\sigma_{\xi}^2$. The system is then highly sensitive to multiplicative noise and there is no stability margin. Obviously in that case, no stationary solution is reachable and the network coherence becomes infinite
at large time. In higher $d>2$, a stability margin appears, the system is then capable to perform the average for small noise. The network coherence is then finite below the noise threshold and for $\sigma^2\geqslant\sigma^2_{{\rm c}}$ the network coherence grows
exponentially again, see Fig~.(\ref{tab2}).  Another quantity that might be extremely insightful in the context of network consensus algorithms is the variance with respect to random initial conditions $u_i(t_0)$ at time $t_0$
\begin{equation}
C_N(t,t_0):= \frac{1}{N}\sum_{i=1}^{N}\mathrm{var}\Big(u_i(t)- \frac{1}{N} \sum_{k=1}^{N}u_k(t_0)  \Big).
\end{equation}
This quantity tells us how the system forgets (or not) about the initial state values on the network and how time correlations grow in the system.
In our formalism, the calculation boils down to compute ${\mathrm G}({\bf r},t,t_0)=\langle u({\bf r},t)u({\bf 0},t_0)\rangle$. Those correlations have not been studied in details in the control literature but are well known in the ageing literature \cite{henkel2008non}.  A problem not addressed in this work is the case of a static (quench) disorder, which may have also a lot of interesting applications in
averaging systems and consensus algorithms. A similar threshold,
known in the context of Anderson localization
\cite{anderson1958absence}, appears in those systems as well, and it
will be interesting to see how localization emerges in distributed
systems with quench disorder. The SHE equation with this type of
disorder is often called the parabolic Anderson problem, see for
example \cite{zel1987intermittency}.
One of the most ambitious perspective would be to show that the noise threshold that we have observed in those classical systems, would also appear in quantum systems. Indeed it is well known in the field of quantum computation \cite{nielsen2010quantum} that noise is the main obstacle to efficient computation, and that above a certain value, computation is no longer possible. This well-known result is called the "quantum threshold theorem" \cite{aharonov1997fault} and it has later been shown that this phenomenon is actually a phase transition \cite{aharonov2000quantum}. It would of great interest if our results could be extended to quantum systems, and prove that a similar threshold theorem would hold for quantum distributed systems and quantum consensus \cite{mazzarella2015consensus}.

This work is partially supported by NSF Awards PHY-1344069 and EECS-1408442. NA is
grateful to Thimoth\'ee Thiery for discussions during the KPZ program
at KITP. CS is grateful to the Labex NEXT, the CSHL, and the MUSE
IDEX Toulouse contract for funding his visit to NY.

\newpage

\bibliographystyle{ieeetr}
\bibliography{biblio}

\end{document}